\documentclass[prd,preprintnumbers,12pt]{revtex4}
\usepackage{epsfig}
\usepackage{graphicx}

\newcommand{\be}{\begin{equation}}
\newcommand{\dd}{\displaystyle}
\newcommand{\ee}{\end{equation}}
\newcommand{\bea}{\begin{eqnarray}}
\newcommand{\eea}{\end{eqnarray}}

\newcommand{\nn}{\nonumber}
\newcommand{\de}{\partial}

\begin{document}
\preprint{{\bf DFF-413/05/04}}
\title{Moose models with vanishing $S$ parameter}

\author{R. Casalbuoni, S. De Curtis, D. Dominici}
\affiliation{Department of Physics, University of Florence, and
INFN, Florence, Italy}
%\author{M. Grazzini}
%\affiliation{CERN, Department of Physics, TH Division, Geneva, Switzerland}\
\date{\today}

\begin{abstract}
\noindent In the linear moose framework, which naturally emerges
in deconstruction models, we show that there is a unique solution
for the vanishing of the $S$ parameter at the lowest order in the
weak interactions. We consider an effective gauge theory based on
$K$ $SU(2)$ gauge groups, $K+1$ chiral fields and electroweak
groups $SU(2)_L$ and $U(1)_Y$ at the ends of the chain of the
moose. $S$ vanishes when a link in the moose chain is cut. As a
consequence one has to introduce a dynamical non local field
connecting the two ends of the moose. Then the model acquires  an
additional custodial symmetry which protects this result. We
examine also the possibility of a strong suppression of $S$
through an exponential behavior of the link couplings as suggested
by Randall Sundrum metric.

\end{abstract}
\pacs{xxxxxxx} \maketitle
\section{Introduction \label{sec:0}}

Even after the very precise measurements made at LEPI, LEPII,
SLC and TEVATRON, the problem of the nature of the electroweak
symmetry breaking remains to be unveiled. In particular, the Higgs
particle has not been observed.

An approach to the problem of electroweak symmetry breaking is
offered by technicolor (TC) theories where  the Higgs is
realized as a composite state of strongly interacting fermions,
the techniquarks. However the TC solution suffers  the drawback
arising from the electroweak precision measurements. These
difficulties, especially the ones coming from the experiments at
the $Z$ pole, can be summarized in a single observable.
This quantity is the so called $S$ parameter
\cite{Peskin:1990zt},\cite{Peskin:1992sw}
or the related $\epsilon_3=g^2 S/(16\pi)$
\cite{Altarelli:1991zd}. The experimental value of $\epsilon_3$ is
of the order  of $10^{-3}$ \cite{Altarelli:1998et}, whereas the
value expected in TC theories is naturally  an order of magnitude bigger. There are other two
important quantities which parameterize the electroweak
observables at the $Z$-pole, $\epsilon_1$ and $\epsilon_2$
\cite{Altarelli:1991zd} (the parameters $T$ and $U$ in the
notations of Ref. \cite{Peskin:1990zt,Peskin:1992sw}). Contrarily to
$\epsilon_3$ these two parameters can be made generally small due
to the custodial symmetry $SU(2)$ which is typically present in
the TC models. As far as $\epsilon_3$ is concerned an enhanced symmetry
$SU(2)\otimes SU(2)$ is necessary to make it small
\cite{Inami:1992rb}. It turns out that producing this symmetry is
quite difficult in TC theories.

A possible solution  to the problem of $\epsilon_3$ was proposed
in Refs. \cite{Casalbuoni:1995yb},\cite{Casalbuoni:1996qt} (see
also Ref. \cite{Appelquist:1999dq}). This was realized in terms of
an effective TC theory  of non linear $\sigma$-model scalars and
massive gauge fields. The model contains three non linear $SU(2)$
fields and two $SU(2)$ gauge groups (before introducing the
electroweak gauge interactions). The physical spectrum consists of
three massless scalar fields (the Goldstone bosons  giving mass to
the gauge vector particles) and two triplets of massive vector
fields degenerate in mass and couplings.  This model, named
degenerate BESS model (D-BESS), has an enhanced custodial symmetry
such to allow $\epsilon_3=0$ at the lowest order in the
electroweak interactions.

A more general  case with $n+1$ gauge groups $SU(2)$ and $n+2$ non
linear $\sigma$-model scalar fields was  studied in Ref.
\cite{Casalbuoni:1989xm} .
 This model has the same content of
fields and symmetries of the open linear moose
\cite{Arkani-Hamed:2001ca,Arkani-Hamed:2001nc,Hill:2000mu,Cheng:2001vd}
but a more general lagrangian.  In fact, in the linear moose
models the scalar fields interact only with their nearest
neighborhood gauge groups along the chain. Therefore a  linear
moose  looks as a linear lattice with lattice sites represented by
the gauge groups and links by the scalar fields. This structure is
particularly interesting and it is the basis of the
"deconstruction" models
\cite{Arkani-Hamed:2001ca,Arkani-Hamed:2001nc,Hill:2000mu,Cheng:2001vd}.
Its  continuum limit leads to a 5-dimensional gauge theory. It is
also possible to start with a 5-dimensional theory, discretize (or
deconstruct) the fifth dimension and obtain a linear moose.

The typical value of $\epsilon_3$ obtained in the linear moose
models is of the same order of magnitude as in the TC theories.
However, in this class of models we have an example, the D-BESS
model, giving $\epsilon_3=0$ (at the lowest order in weak
interactions). Then  it seems natural to investigate  the possible
solutions to $\epsilon_3=0$ within the moose models. We have
indeed found a general solution which turns out to be a simple
generalization of the mechanism present in  D-BESS.

In Section \ref{linear} we introduce the notations and the main
constitutive elements of a linear moose model such to describe the
electroweak symmetry breaking in a minimal way. In particular this
requires that, after the gauge fields have acquired mass, only
three massless scalar fields (the ones giving masses to $W$ and
$Z$) should remain in the spectrum, and also that the fermions
couple to the electroweak gauge fields in the standard way. In
this case there is no contribution to $\epsilon_3$ from fermions.

In Section \ref{section3} we make use of the analysis of Ref.
\cite{Peskin:1992sw} to get a general expression for $\epsilon_3$.
The result can be written in a very compact form in terms of a
particular matrix element (the one between the ends of the moose)
of $M_2^{-2}$, where $M_2$ is the quadratic mass matrix of the
gauge bosons. We express also this matrix element
 in terms of the decay coupling constants (or link couplings)
of the scalar fields. As a by-product we get the result that
$\epsilon_3$ is a semi-positive definite  expression (see also
Refs. \cite{Barbieri:2003pr,Chivukula:2004kg,Hirn:2004ze}).

In Section \ref{cut} we investigate the possible models with
$\epsilon_3=0$. We show that the unique solution corresponds to
have a vanishing decay coupling constant (or more, in an
independent way). However, letting two or more couplings to zero
in a correlated way leads to a non vanishing $\epsilon_3$. This
solution corresponds to cut a link and to disconnect the linear
moose in two parts. By choosing this option one needs to introduce
an additional scalar field in order to have the right number of
degrees of freedom to give masses to $Z$ and $W$. This new
dynamical field is provided by
 a non local field (in
lattice space), connecting the two  ends of the original moose.
Therefore, at the lowest order in the weak interactions, the
original moose splits in three disconnected parts producing an
enhancement of the custodial symmetry from $SU(2)$ to
$SU(2)\otimes SU(2)$ and leading to a vanishing $\epsilon_3$. We
have also examined the case of a linear moose with a reflection
symmetry with respect to the ends of the moose. It is again
possible to have $\epsilon_3=0$ but only for an even number of
gauge groups. The original D-BESS model corresponds exactly to
this latter case with two gauge groups. Another relevant aspect of
cutting a link is that the Goldstone bosons related to the weak
symmetry breaking are associated only to the non local field. As a
consequence the unitarity properties of these models are the same
as in the  Higgsless Standard Model.

In Section \ref{bess} we give a detailed description of the D-BESS
model showing its relation to the linear moose case with a cut.

A value of $\epsilon_3$ strongly suppressed is equally acceptable
as the case $\epsilon_3=0$. Therefore, in Section
\ref{section:VIII}, we examine the possibility of substituting the
cut of a link with a strong suppression of the corresponding
coupling. In particular we have examined the possibility of an
exponential law for the couplings. In this way the decay constant
at one of the ends of the moose is exponentially suppressed with
respect to all the others. As a result $\epsilon_3$ is strongly
suppressed in agreement with our findings in Section \ref{cut}. We
have also examined a power-like behavior of the couplings with
similar results. By requiring reflection symmetry in both the
previous cases we have shown that the suppression is present only
for an even number of gauge groups.  At the end of this Section we
have studied the continuum limit of the linear moose with
exponential law, which corresponds to a 5-dimensional gauge theory
with a Randall Sundrum metric \cite{Randall:1999vf}. Again we find
a suppression, although not as large as in the discrete case.

In Section \ref{section:VI}   we study the possibility of
extending the linear moose to a planar one. In particular  we show
that no loops are allowed on the plane and that, by a convenient
redefinition of the gauge couplings, the expression for
$\epsilon_3$ is the same as in the linear case.

Conclusions are given in Section \ref{conclusions}.

The Appendix \ref{appendixA} is devoted to the explicit
calculation of $\epsilon_3$ for the linear and the planar moose.
In Appendix \ref{appendixB} we prove the main result of Section
\ref{cut}.

\section{A linear moose model for the electroweak symmetry breaking}
\label{linear}

Following   the idea of the dimensional deconstruction
\cite{Arkani-Hamed:2001ca,Arkani-Hamed:2001nc,Hill:2000mu,Cheng:2001vd}
 and the hidden gauge symmetry
 approach applied  to the
strong interactions
\cite{Bando:1985ej,Bando:1988ym,Bando:1988br,Son:2003et,Hirn:2004ze}
 and
to the electroweak symmetry breaking
\cite{Casalbuoni:1985kq,Casalbuoni:1989xm,Hirn:2004ze},  we consider
$K+1$ non linear $\sigma$-model scalar fields $\Sigma_i$, ${i=1,\cdots ,K+1}$,
$K$ gauge groups, $G_i$, ${i=1,\cdots ,K}$
 and a global symmetry $G_L\otimes
G_R$. Since the aim of this paper is  to investigate a minimal
model of electroweak symmetry breaking,
 we will assume $G_i=SU(2)$,
$G_L\otimes G_R=SU(2)_L\otimes SU(2)_R$. The Standard Model (SM)
gauge group
  $SU(2)_L\times U(1)_Y$ is obtained by gauging
a subgroup of $G_L\otimes G_R$. The $\Sigma_i$ fields can be
parameterized as $\Sigma_i=\exp{(i/(2f_i)\vec \pi_i\cdot \vec
\tau})$ where $\vec \tau$ are the Pauli matrices and $f_i$ are
$K+1$ constants that we will call link couplings.

The transformation properties of the fields are
 \bea
&&\Sigma_1\to L\Sigma_1 U_1^\dagger,\nn\\
&&\Sigma_i\to U_{i-1}\Sigma_i U_i^\dagger\
,\,\,\,\,\,\,\,i=2,\cdots,K,
\nn\\
&&\Sigma_{K+1}\to U_K\Sigma_{K+1} R^\dagger, \eea with $U_i\in
G_i$, $i=1,\cdots,K$, $L\in G_L$, $R\in G_R$.

The lagrangian   is given by \be {\cal
L}=\sum_{i=1}^{K+1}f_i^2{\rm Tr}[D_\mu\Sigma_i^\dagger
D^\mu\Sigma_i]-\frac 1 2\sum_{i=1}^K{\rm Tr}[(F_{\mu\nu}^i)^2],
\label{lagrangian:l} \ee with  the covariant derivatives  defined
as follows \bea &D_\mu\Sigma_1=\de_\mu\Sigma_1+i\Sigma_1 g_1
A_\mu^1,&\nn\\
&D_\mu\Sigma_i=\de_\mu\Sigma_i-ig_{i-1}A_\mu^{i-1}\Sigma_i+i\Sigma_i
g_i A_\mu^i,&\,\,\,\,\,\,\,i=2,\cdots,K,\nn\\
&D_\mu\Sigma_{K+1}=\de_\mu\Sigma_{K+1}-ig_{K}A_\mu^{K}\Sigma_{K+1},&\eea
where $A_\mu^i$ and $g_i$ are the gauge fields and gauge coupling
constants associated to the groups $G_i$, $i=1,\cdots ,K$.

\begin{figure}[h] \centerline{
\epsfxsize=12cm\epsfbox{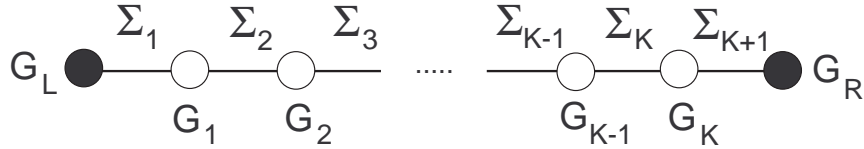} } \caption {{\it  The linear
moose model. \label{fig:2} }}
\end{figure}

The model described by the
lagrangian (\ref{lagrangian:l})
is represented in Fig. \ref{fig:2}. Notice that the
field defined as
\be U=\Sigma_1\Sigma_2\cdots\Sigma_{K+1}
\label{chiral}\ee
is the usual chiral field: in fact
it transforms as
$
U\rightarrow LUR^\dagger
$
and it is invariant under the $G_i$ transformations.

The mass matrix of the gauge fields can be obtained by choosing
$\Sigma_i=I$ in Eq. (\ref{lagrangian:l}). We find \be {\cal
L}_{\rm mass}=\sum_{i=1}^{K+1}f_i^2{\rm Tr}[(g_{i-1}A_\mu^i-g_i
A_\mu^i)^2]\equiv \frac 1 2\sum_{i,j=1}^{K+1}(M_2)_{ij}A_\mu^i
A^{\mu j}, \label{lmass}\ee with \be
(M_2)_{ij}=g_i^2(f_i^2+f_{i+1}^2)\delta_{i,j}-g_i g_{i+1}f_{i+1}^2
\delta_{i,j-1}-g_j g_{j+1}f_{j+1}^2  \delta_{i,j+1}. \label{m2}\ee
The squared mass matrix can be diagonalized through an orthogonal
transformation $S$. By calling  $\tilde A_\mu^n$, $n=1,\cdots,K$
the mass eigenstates, and $m^2_n$ the squared mass eigenvalues, we
have \be A_\mu^i=\sum_{n=1}^KS^i_n\tilde A_\mu^n,\ee and \be
S^i_m(M_2)_{ij}S^j_n=m^2_n\delta_{m,n}.\ee We will assume $m_n\neq
0$, otherwise the model describes an unphysical situation.

The vector meson decay constants are defined in terms of the
matrix elements of the vector and axial vector currents between
the vacuum and the one vector meson states, i.e. \bea \langle
0|J_{V\mu}^a|\tilde
A^n_b(p,\epsilon)\rangle&=&g_{nV}\delta^{ab}\epsilon_\mu,\nn\\
0|J_{A\mu}^a|\tilde
A^n_b(p,\epsilon)\rangle&=&g_{nA}\delta^{ab}\epsilon_\mu,\eea
where $\vert \tilde A^n_b(p,\epsilon)\rangle$ is the $b$ component
of the single particle state of the $n$-vector boson with
polarization $\epsilon_\mu$. Notice that the vector and axial
vector currents are defined as the conserved currents associated
to the global symmetry $G_L\otimes G_R$ acting at the ends of the
moose. Therefore the vector meson decay constants can be very
easily obtained by considering the contribution of the vector
mesons to the canonical currents. Notice that
 only the scalar fields
$\Sigma_1$ and $\Sigma_{K+1}$ transform under the vector and axial
transformations according to
 \bea {\rm vector}&:&~~~~\Sigma_1\to
T\Sigma_1,~~~\Sigma_{K+1}\to \Sigma_{K+1}T^\dagger,\nn\\ {\rm
axial}&:&~~~~\Sigma_1\to V\Sigma_1,~~~\Sigma_{K+1}\to
\Sigma_{K+1}V.\eea Then, the contributions of the vector mesons to
the conserved vector and axial vector currents are  \bea
J_{V\mu}^a\Big|_{\rm vector~mesons}&=&f_1^2g_1A_{\mu}^{1a}+f_{K+1}^2g_KA_\mu^{Ka},\nn\\
J_{A\mu}^a\Big|_{\rm vector~
mesons}&=&f_1^2g_1A_{\mu}^{1a}-f_{K+1}^2g_KA_\mu^{Ka}.\eea It
follows \bea g_{nV}&=&f_1^2g_1S^1_n+f_{K+1}^2g_KS^K_n,\nn\\
g_{nA}&=&f_1^2g_1S^1_n-f_{K+1}^2g_KS^K_n.\label{couplings} \eea

\section{Determination of $\epsilon_3$}
\label{section3}

To compute the new physics contribution to the electroweak
parameter $\epsilon_3$ \cite{Altarelli:1991zd} we will make use of
the dispersive representation given in Refs.
\cite{Peskin:1990zt,Peskin:1992sw} for the related parameter $S$
($\epsilon_3=g^2 S/(16\pi)$, where $g$ is the $SU(2)_L$ gauge
coupling) \be
\epsilon_3=-\frac{g^2}{4\pi}\int_0^\infty\frac{ds}{s^2}
Im\left[\Pi_{VV}(s)-\Pi_{AA}(s)\right],\ee where $\Pi_{VV}(AA)$ is
the current-current correlator \be \int d^4x e^{-iq\cdot x}\langle
J^\mu_{V(A)}J^\nu_{V(A)}\rangle=ig^{\mu\nu}\Pi_{VV(AA)}(q^2)+(q^\mu
q^\nu~{\rm terms}).\ee It should be noticed that the $\epsilon_3$
parameter is evaluated with reference to the SM, and therefore the
corresponding contributions should be subtracted. For instance the
contribution of the pion pole to $\Pi_{AA}$, that is of the
Goldstone particles giving mass to the $W$ and $Z$ gauge bosons,
does not appear in $\epsilon_3$. In the  model described by the
lagrangian (\ref{lagrangian:l}) all the new physics contribution
comes from the new vector bosons (we are assuming the standard
couplings for the fermions to $SU(2)_L\times U(1)_Y$). Therefore
from\be
Im\,\Pi_{VV(AA)}=-\pi\sum_{Vn,An}g^2_{nV,nA}\delta(s-m^2_n),\ee we
get \be
\epsilon_3=\frac{g^2}4\sum_n\left(\frac{g_{nV}^2}{m^4_n}-\frac{g^2_{nA}}{m^4_n}\right).\ee
Substituting the expressions (\ref{couplings}) for the decay
vector couplings we find \be \epsilon_3=g^2g_1 g_K f_1^2
f_{K+1}^2\sum_n\frac{S^1_n S^K_n}{m^4_n}=g^2g_1 g_K f_1^2
f_{K+1}^2(M_2^{-2})_{1K}.\label{eps3} \ee In Appendix
\ref{appendixA} we have derived the following explicit expression
for $\epsilon_3$, valid for a generic linear moose model (the same
result has been obtained  in \cite{Hirn:2004ze}): \be \epsilon_3
=g^2\sum_{i=1}^K\frac{(1-y_i)y_i}{g_i^2}, \label{eps_lin} \ee
where we have introduced the following notations
 \be y_i=\sum_{j=1}^i
x_j,~~~~~~~x_i=\frac{f^2}{f_i^2},~~~~~i=1,\cdots,K+1,
\label{xi}\ee with \be \frac 1 {f^2}=\sum_{i=1}^{K+1}\frac 1
{f_i^2}. \label{f2}\ee Therefore $\sum_{i=1}^{K+1}x_i=1$.

From Eq. (\ref{eps_lin}) it follows that for an open moose one has
always \be \epsilon_3\ge 0,\ee since $0\leq y_i\leq 1$,
$i=1,\cdots,K+ 1$. The positivity of $\epsilon_3$ is a simple
consequence of the positivity of all the matrix elements of
$M_2^{-1}$. This can be proved by using the decomposition of $M_2$
(see Eq. (\ref{m2})) in  triangular matrices. The positivity of
$\epsilon_3$  was already noticed \cite{Barbieri:2003pr} for  the
warped 5 dimensional  models (whose deconstruction generates
linear moose models) and for the deconstructed QCD
\cite{Chivukula:2004kg}.

Furthermore if all the  $f_i$ and the gauge couplings $g_i$ are of
the same order of magnitude, the typical size  for $\epsilon_3$ is
\be \epsilon_3\sim \frac{g^2}{g_i^2}.\ee However, since the
experimental value of $\epsilon_3$ is of the order  $10^{-3}$
\cite{Altarelli:1998et}, in order to get a realistic model, one
should have strongly coupled vector bosons $A_\mu^i$.

As a simple example, let us consider the case $K=2$.
The result for $\epsilon_3$ is:
 \be \epsilon_3=
\frac{g^2}{g_1^2
g_2^2}f_1^2f_2^2f_3^2\frac{(f_1^2+f_2^2)g_1^2+(f_2^2+f_3^2)g_2^2}
{(f_1^2f_2^2+f_1^2f_3^2+f_2^2f_3^2)^2}.\label{eq:23}\ee
Analogously for $K=3$ we obtain \bea&&\epsilon_3= \frac{g^2}{g_1^2
g_2^2g_3^2}f_1^2f_2^2f_3^2f_4^2\times\nn\\
&&\frac{(f_1^2f_2^2+f_2^2f_3^2+f_1^2f_3^2)g_1^2g_2^2
+(f_1^2+f_2^2)(f_3^2+f_4^2)g_1^2g_3^2+(f_3^2f_4^2+f_2^2f_3^2+f_2^2f_4^2)g_2^2g_3^2}
{(f_1^2f_2^2f_3^2+f_1^2f_3^2f_4^2+f_2^2f_3^2f_4^2+f_1^2f_2^2f_4^2)^2}.\label{eq:24}\eea

\section{Cutting a link}
\label{cut}

Is there a possibility to get $\epsilon_3=0$ at  the lowest order
in the weak interactions? This can be realized by noticing that
if one  of the $f_i$, with $i=2,\cdots ,K$,
 vanishes, the mass matrix
$M_2$ is block-diagonal. The case $f_1=0$ or $f_{K+1}=0$ implies
the vanishing of $\epsilon_3$ in a trivial way due to Eq.
(\ref{eps3}) and the fact that the matrix $M_2$ is not singular
under these hypothesis. This general result can  be explicitly
verified for $K=2$ and $K=3$ (see Eqs. (\ref{eq:23}) and
(\ref{eq:24})). We will refer to this situation as "cutting a
link". In such a case also $M_2^{-2}$ is block-diagonal, implying
the vanishing of $\epsilon_3$. This can be also derived from   the
explicit expression (\ref{eps_lin}). Let us choose $f_{m}=0$, then
$x_i=\delta_{i, m}$ and $y_i=\sum_{j=1}^i \delta_{j, m}=\theta_{i,
m}$, where we have defined the discrete step function \be
\theta_{i,j}=\Big\{^{\dd{1,~~~{\rm for}~ i\ge j,}}_{\dd{0,~~~{\rm
for}~ i<j.}}\label{eq:25}\ee Then we obtain \be \epsilon_3= g^2
\sum_{i=1}^K \frac {(1-\theta_{i, m})\theta_{i, m}}{g_i^2}=0.
\label{step} \ee

 However cutting a link corresponds
to lose one  scalar multiplet which is necessary to give masses to
the gauge bosons of the standard $SU(2)_L\times U(1)_Y$. We can
solve this problem by adding to the lagrangian of the linear moose
a term given by \be f_0^2 Tr[
\partial_\mu U^\dagger \partial^\mu U],
\ee where $U$ is the chiral field given in Eq. (\ref{chiral}) and
$f_0$ is a new parameter related to the Fermi scale.

\begin{figure}[h] \centerline{
\epsfxsize=15cm\epsfbox{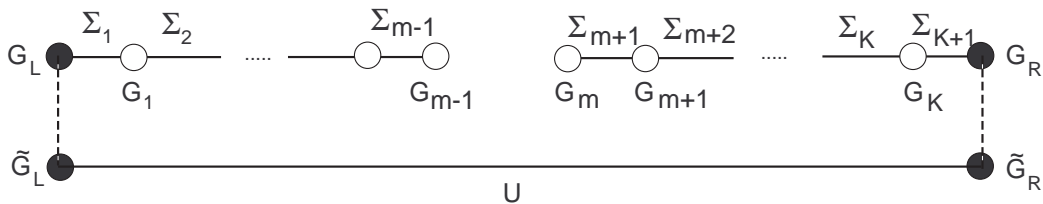} } \caption {{ \it Graphic
representation of the linear moose model with the $m$  link cut
described by the lagrangian (\ref{lagrangian:c}).  The dashed
lines represent the identification of the global symmetry groups
after weak gauging. \label{fig:5} }}
\end{figure}

Correspondingly there is an enhancement of the symmetry from
$G_L\otimes G_R\otimes\prod_{i=1}^{K} G_i$ to $G_L\otimes
G_R\otimes\tilde G_L\otimes \tilde G_R\otimes\prod_{i=1}^{K} G_i$,
where $\tilde G_{L(R)}$ is a copy of $G_{L(R)}$ and $U$ transforms
as \be U\rightarrow \tilde L U \tilde R^\dagger, \ee with $\tilde
L(\tilde R)\in \tilde G_{L(R)}$. The lagrangian for the model,
with the $m$ link  cut, is given by \bea {\cal L}&=& f_0^2
Tr[\partial_\mu U^\dagger \partial^\mu
U]+\sum_{i=1}^{m-1}f_i^2{\rm
Tr}[D_\mu\Sigma_i^\dagger D^\mu\Sigma_i]\nn\\
&+& \sum_{i=m+1}^{K+1}f_i^2{\rm Tr}[D_\mu\Sigma_i^\dagger
D^\mu\Sigma_i]-\frac 1 2\sum_{i=1}^K{\rm Tr}[(F_{\mu\nu}^i)^2].
\label{lagrangian:c} \eea As already mentioned, it has an enhanced
symmetry with respect to the lagrangian (\ref{lagrangian:l}) since
the global symmetry  $\tilde G_L\otimes \tilde G_R$ under which
the kinetic term for the field $U$ is invariant does not coincide
with the symmetry $G_L\otimes G_R$ acting upon the scalar fields
$\Sigma_1$ and $\Sigma_{K+1}$.
 These two
global symmetries are to be identified only after the gauging of
the  electroweak symmetry. The  model  corresponding to the
lagrangian (\ref{lagrangian:c}) is shown in   Fig. \ref{fig:5}.
Before the weak gauging
 we have three disconnected chains and this is the
reason why the symmetry gets enhanced. Clearly the main
difference with  respect to the linear moose
  model is the fact
that a  link is cut and the invariant term containing the  scalar
field $\Sigma_{m}$ is substituted by the invariant involving the
field $U$ coupling the two ends of the chain. Cutting a link
implies that, in the unitary gauge, the gauge fields $A_\mu^i$
become massive by eating the $\Sigma_i$ fields, while the $U$
field contains the Goldstone bosons which  give masses to the
standard gauge bosons once the gauge group $SU(2)_L\times U(1)_Y$
is switched on. This additional term does not contribute to
$\epsilon_3$ because the $U$ field does not couple to the gauge
fields $A_\mu^i$, $i=1,\cdots ,K$; as a consequence the gauge
boson mass matrix $M_2$ remains unchanged.

It is also worth to notice that the enhanced
symmetry acts as a custodial symmetry and this explains why the
parameter $\epsilon_3$ is vanishing \cite{Inami:1992rb}.

Of course the enhanced symmetry is broken by the weak gauging and
corrections to $\epsilon_3$ of order $\alpha(M_Z/M)^2$, where $M$
is the mass scale of the new vector bosons (see Refs.
\cite{Casalbuoni:1995yb,Casalbuoni:1996qt}), are expected.

In the linear moose model described by the lagrangian
(\ref{lagrangian:l}) one has the possibility of making
$\epsilon_3$ small by choosing one $f_i$ much smaller than the
other ones: an explicit calculation will be
presented in Section \ref{section:VIII}. However in this case there is no additional symmetry
which protects the result.

In both cases, the parameters $\epsilon_1$ and $\epsilon_2$ are
zero, at the lowest order in the weak interactions, because of
the presence of the usual $SU(2)_{L+R}$ custodial symmetry.

In the Appendix B of \cite{Casalbuoni:1989xm} it was already shown
that in the case of $G_i=SU(2)$ for $i=1,\cdots ,K$
 and $G_{L(R)}=SU(2)_{L(R)}$ one exactly gets
$\epsilon_1=0$ and, requiring the decoupling of the gauge fields
$A_\mu^i$, the parameter $f_0$ in Eq. (\ref{lagrangian:c})
satisfies $f_0^2=(\sqrt{2} G_F)^{-1}$.

Concerning the fermions, if we assume the usual representation
assignments with respect to $SU(2)_L\times U(1)_Y$, mass terms can
be generated by Yukawa couplings to
 the $U$ field. In this case fermion couplings to $W$ and $Z$ are
the standard ones if we neglect the effect of the mixing with the
additional vector bosons. Of course it would be possible to add
new couplings of the fermions to the gauge bosons. These new
couplings would modify $\epsilon_3$ but, in order to get the
necessary cancellation to fulfill the electroweak constraints, one
would need a fine tuning of the parameters (as an example, see the
BESS model corresponding to $K=1$, Ref. \cite{Casalbuoni:1985kq}).

\begin{figure}[h] \centerline{
\epsfxsize=12cm\epsfbox{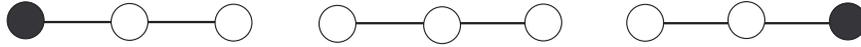} } \caption {{\it  For $K$
odd, putting one of the $f_i$'s to zero in a reflection invariant
model, one is left with a string containing more vector fields
than scalars. \label{fig:3} }}
\end{figure}

Up to now we have not required the reflection invariance with
respect to the ends of the moose. If we do require  invariance we
get the following relations among the couplings \be
f_i=f_{K+2-i},~~~g_i=g_{K+1-i}.\ee If $K$ is odd we have an even
number of scalar fields and, putting one link coupling $f_i$ to
zero, implies to cut two links (the two connected by the
reflection symmetry). This  leads to an unphysical situation,
since  a multiplet of vector fields  remains massless. This is
illustrated in Fig. \ref{fig:3}. The original string is broken in
three pieces with the central one containing more vector fields
than scalar ones. As a consequence there are massless vector
fields in the spectrum of the theory. In this case the matrix
$M_2$ is singular and the Eq. (\ref{eps3}) is not applicable as it
stands. However, as we shall see, $\epsilon_3$ can be defined
through a limiting procedure.

\begin{figure}[h] \centerline{
\epsfxsize=8cm\epsfbox{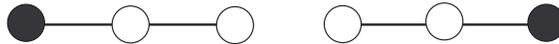} } \caption {{\it  For $K$
even, cutting the central link we are left with two strings, each
of them ending with a gauge field.  \label{fig:4} }}
\end{figure}
The situation is different for $K$ even, since in this case we can
cut the central link, remaining as depicted in Fig. \ref{fig:4}.
That is we are left with two strings disconnected, each of them
with a gauge field at one end point. Therefore for $K=2N$ the
lagrangian is given by the formula (\ref{lagrangian:c}) with
$m=N+1$ with the field $U$ expressed as \be
U=\Sigma_1\cdots\Sigma_{N+1}\cdots\Sigma_{2N+1}. \ee

Another interesting point is that, due to the reflection
invariance, the mass matrices of the two disconnected strings
containing the $\Sigma$ fields are equal. Therefore there is
complete degeneracy between vector and axial vector resonances.
The models so obtained can be considered as a generalization of
the D-BESS model, corresponding to $N=1$, as shown in Section
\ref{bess}.

As a general result,  it is possible
to build a model with $\epsilon_3=0$ and an extra custodial symmetry
even without requiring the reflection invariance.

Finally let us mention the unitarity bounds. In general for a cut
linear moose model the longitudinal components of the electroweak
gauge bosons are only  coupled to the $U$ field. As a consequence
the corresponding scattering amplitudes violate partial wave
unitarity at the same energy scale as in the Higgsless SM.
Therefore the violation of unitarity is not postponed to higher
scales as in the 5 dimension Higgsless  model, which, however,
seem to be  difficult to be reconciled with the precision
electroweak measurements
 unless one includes brane kinetic terms
 \cite{Barbieri:2003pr,Cacciapaglia:2004jz,Davoudiasl:2004pw}.

\section{The D-BESS Model}
\label{bess}

From the general formalism developed in the previous Sections,
 assuming $K=2$ and  reflection invariance,
 one can easily
derive the   lagrangian of Ref. \cite{Casalbuoni:1989xm},
describing new vector and axial vector gauge bosons in the
Higgsless SM, in two particular cases. Let us recall that,
requiring gauge invariance and symmetry under reflection, the most
general invariant lagrangian is \be {\cal L}=-\frac 1 4 v^2\left[
a_1 I_1+a_2 I_2+a_3 I_3 +a_4 I_4\right]-\frac 1 2\sum_{i=1}^2 {\rm
Tr}[(F_{\mu\nu}^i)^2]\label{lbess}\ee with \bea &I_1={\rm
Tr}[(V_1-V_2-V_3)^2],~~~I_2={\rm
Tr}[(V_1+V_3)^2],&\nn\\
&I_3={\rm Tr}[(V_1-V_3)^2],~~~I_4={\rm Tr}[V_2^2],& \eea where
 \be V_1^\mu=\Sigma_1^\dagger D^\mu\Sigma_1,~~~
V_2^\mu=\Sigma_2
D^\mu\Sigma_2^\dagger,~~~V_3^\mu=\Sigma_2(\Sigma_3
D^\mu\Sigma_3^\dagger)\Sigma_2^\dagger,\ee and \bea
&D_\mu\Sigma_1=\de_\mu\Sigma_1+i\Sigma_1 g_1 A_\mu^1,&\nn\\
&D_\mu\Sigma_2=\de_\mu\Sigma_2-i g_1 A_\mu^1\Sigma_2+i\Sigma_2 g_2
A_\mu^2,\nn\\
&D_\mu\Sigma_3=\de_\mu\Sigma_3-i g_2 A_\mu^2\Sigma_3.& \eea The
invariance under reflections implies
 \be
\Sigma_3\leftrightarrow\Sigma_1^\dagger,~~~
\Sigma_2\leftrightarrow\Sigma_2^\dagger~~~A^1_\mu \leftrightarrow
A^2_\mu ~~~g_1=g_2, \ee where $A^1_\mu$ and $A^2_\mu$ are the
gauge fields related to the gauge groups $G_1$ and $G_2$
respectively.

\begin{figure}[h] \centerline{
\epsfxsize=12cm\epsfbox{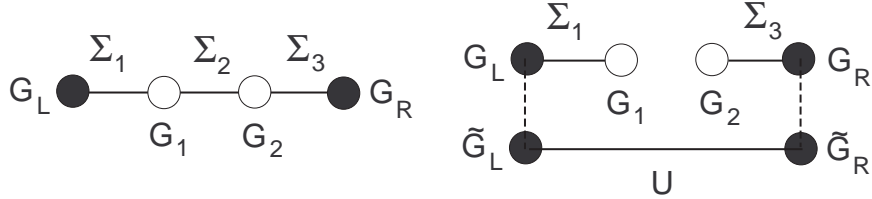} } \caption {{ \it The left
panel gives a graphic representation of the  lagrangian
(\ref{lbess}) for $a_1=0$, $a_2=a_3$. The right panel gives a
graphic representation of the D-BESS model lagrangian
(\ref{lagrangian:2}). The dash lines represent the identification
of the global symmetry groups after the electroweak
gauging.\label{fig:1} }}
\end{figure}

We can now select two particular cases:\hfill\\\\\noindent 1) -
\underline{The linear moose model}. By choosing \be
a_1=0,~~~a_2=a_3,\ee we have \be{\cal L}=\sum_{i=1}^3 f_i^2{\rm
Tr}[D_\mu\Sigma_i^\dagger D^\mu\Sigma_i]-\frac 1 2\sum_{i=1}^2
{\rm Tr}[(F_{\mu\nu}^i)^2], \label{lagrangian:b1} \ee with \be
f_1^2=f_3^2=\frac 1 2 a_2 v^2,~~~ f_2^2=\frac 1 4 a_4 v^2.\ee This
is indeed the lagrangian for a linear moose with three links and
two gauge fields with reflection invariance (see Refs.
\cite{Georgi:1990xy} and \cite{Son:2003et}). The corresponding
diagram is shown in the left panel of Fig. \ref{fig:1}.

\noindent 2) - \underline{The D-BESS model
\cite{Casalbuoni:1995yb,Casalbuoni:1996qt}}. This corresponds to
the choice \be a_4=0,~~~a_2=a_3, \label{condition:1}\ee giving \be
{\cal L}_{\rm D-BESS} \label{lagrangian:2}=f^2{\rm Tr}[\de_\mu
U^\dagger\de^\mu U]+ f_1^2\left({\rm Tr}[D_\mu\Sigma_1^\dagger
D^\mu\Sigma_1]+{\rm Tr}[D_\mu\Sigma_3^\dagger
D^\mu\Sigma_3]\right), \ee with \be f^2=\frac 1 4 a_1 v^2,~~~
f_1^2=\frac 1 2 a_2 v^2,\ee and \be U=\Sigma_1\Sigma_2\Sigma_3.\ee
 The  diagram corresponding to the previous
lagrangian is shown in the right panel of Fig. \ref{fig:1}. Before
the electroweak gauging we have three disconnected chains and this
is the reason why the symmetry $SU(2)_L\otimes\prod_{i=1}^2
SU(2)_i\otimes SU(2)_R$ gets enhanced to $[SU(2)\otimes SU(2)]^3$.

 We have shown in
\cite{Casalbuoni:1995yb} that in order to have vanishing parameter
$\epsilon_3$, or $S$, at the lowest order in the weak
interactions, $a_4=0$ is necessary. This is equivalent to
eliminate from the lagrangian the term corresponding to the
central link. The requirement $a_2=a_3$ implies degeneracy between
vector and axial vector gauge bosons. Since the contribution of
the vector and  of the axial vector particles are of opposite
sign, one gets exactly $\epsilon_3=S=0$ at the leading order.
However, as we have already noticed in Section \ref{section3}, it
is possible to build a model with $\epsilon_3=0$ and an extra
custodial symmetry even without requiring the reflection
invariance. In other words the degeneracy of vector and axial
vector resonances is not necessary to ensure $\epsilon_3=0$.

\section{Sewing the cut}
\label{section:VIII}

We have shown in Appendix \ref{appendixB} that, in order to get a
vanishing $\epsilon_3$, the necessary and sufficient condition is
that one and only one of the
  link couplings $f_i$ is zero. As a consequence, by
requiring reflection invariance, $\epsilon_3=0$ can be achieved
only if $K$ is even. On this basis it is easy to see how a
suppression of $\epsilon_3$ (a smoother situation with respect to
the vanishing) can be realized. In fact, it will be enough to
require that a link is suppressed with respect to all the others.
In this case however it is not necessary to consider the
additional dynamical degree of freedom given by the chiral field
$U$, and therefore there is no additional custodial
 symmetry
for $\epsilon_3$.

A simple model grasping the main features is obtained by assuming
an exponential law for the  link couplings $f_i$, and equal gauge
couplings \be f_i=\bar fe^{c(i-1)}, ~~~~~g_i=\tilde
g~~~i=1,\cdots,K+1 \label{f-exp}.\ee

\begin{figure}[h] \centerline{
\epsfxsize=16cm\epsfbox{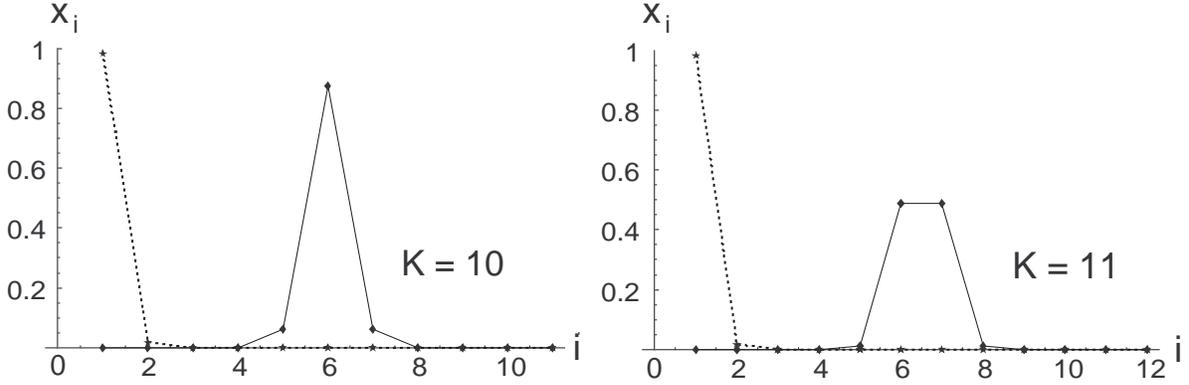} } \caption {{ \it The
behavior of $x_i=f^2/f_i^2$ vs. $i$ for  $c=2$,
 $K=10$ (left panel)  and
$K=11$ (right panel). The dotted (continuous) lines correspond to
the choice of link couplings $f_i$  without   reflection symmetry,
given in Eq. (\ref{f-exp}) (with reflection symmetry, given in Eq.
(\ref{f-cosh})).
 \label{xi_exp}}}
\end{figure}

Here $\bar f$ is an overall scale not playing any
role in the dimensionless quantity $\epsilon_3$. By contrast
the relevant
 parameter  is $c$ since it controls the amount
of suppression. By using Eqs.(\ref{eps_lin}), (\ref{xi}) and
(\ref{f2}), we easily obtain \be f^2=\bar f^2e^{cK}\frac{\sinh
c}{\sinh c(K+1)},\ee \be x_i=e^{-2ci}e^{c(K+2)}\frac{\sinh
c}{\sinh c(K+1)},\ee and \be\epsilon_3=\frac 1 4 \left(\frac
g{\tilde g}\right)^2\frac{\sinh(2c(K+1))-(K+1)\sinh 2c}{\sinh
2c\sinh^2(c(K+1))}.\label{eq:46}\ee For increasing $c$, the first
link $f_1$ is more and more suppressed with respect to the other
links. In fact for large $c$  we get \be \epsilon_3\sim
\left(\frac g{\tilde g}\right)^2 e^{-2c}. \ee Therefore the
suppression factor is about $2\times 10^{-2}$ for $c\approx 2$. It
is interesting to look at the behavior of the variables $x_i$ vs.
$i$ for fixed $c$. This is plotted in Fig. \ref{xi_exp}.

\begin{figure}[h] \centerline{
\epsfxsize=16cm\epsfbox{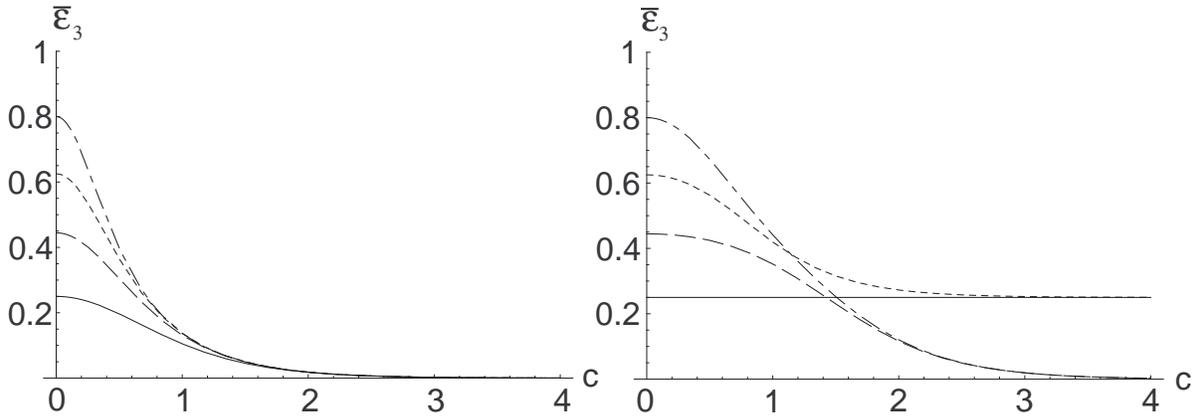} } \caption {{ \it The
behavior of $\bar\epsilon_3=\epsilon_3/(g/\tilde g)^2$ vs. $c$ for
different values of $K$, on the left (right)  panel for the choice
of link couplings $f_i$ without   reflection symmetry, given in
Eq. (\ref{f-exp}) (with reflection symmetry, given in Eq.
(\ref{f-cosh})). The continuous lines correspond to $K=1$, the
dash lines to $K=2$, the dotted lines to $K=3$ and the dash-dotted
lines to $K=4$. \label{exp_all}}}
\end{figure}

From Fig. \ref{xi_exp} we see that for $c=2$ we are practically in
the ideal situation $x_1=1$ and $x_i=0$ for $i\not= 1$. In Fig.
\ref{exp_all} we show the suppression factor in $\epsilon_3$ as a
function of $c$. We see that, in agreement with the analytical
result, $\epsilon_3$ does not depend on $K$ as soon as $c\approx
2$.

We have also considered the case of reflection symmetry by
assuming the link couplings $f_i$ of the form \be f_i={\hat
f}\cosh {[c(1+\frac K 2-i)]} \label{f-cosh}, \ee with $\hat f=\bar
f$ ($\hat f=\bar f/\cosh{(c/2)}$) for $K$ even (odd). With this
choice the central link couplings are equal to $\bar f$. From Fig.
\ref{xi_exp} it appears clearly the difference between $K$ even
and $K$ odd. In particular, for $K$ odd there are two central
$x_i$ much bigger than the others. Therefore, in agreement with
the discussion in Appendix \ref{appendixB}, we expect no
suppression factor. In fact,  for large $c$ the limiting value of
the two central $x_i$ is 0.5 and \be \epsilon_3\to \frac 1
4\left(\frac g{\tilde g}\right)^2.\label{eq:49}\ee The numerical
results for $\epsilon_3$ vs. $c$ are given in Fig. \ref{exp_all}.
We see that the suppression factor is operating only for $K$ even.

We have also analyzed the case of a power-like behavior of the
link couplings \be f_i=\bar f\, i^c,\ee with the related
reflection invariant case.

The results are similar to the exponential case. In order to have
a suppression factor of order $2\times 10^{-2}$ we need $c\approx
3$ for the non symmetric case. On the other hand, when reflection
invariance is required,   we have a similar suppression only for
$K$ even and $c\approx 7.5$. The last result follows from the
$x_i$ distribution which is broader around the central links.
Also in this case there is no suppression for $K$ odd and
$\epsilon_3$ goes to the  limiting value given in Eq.
(\ref{eq:49}).

It is interesting to observe that for any of the previous choices
of $f_i$ we have \be \lim_{c\to 0} f_i=\bar f.\ee Therefore  from
our general expression for $\epsilon_3$ (see Eq.(\ref{eps_lin})),
as well from Eq. (\ref{eq:46}) for $c\to 0$, \be \epsilon_3=\frac
1 6 \left(\frac g{\tilde g}\right)^2\frac{K(K+2)}{K+1},\ee which
coincides with the result of Refs. \cite{Hirn:2004ze} and
\cite{Foadi:2003xa}.

Another interesting aspect is the continuum limit.
It is known that the discretization of a gauge theory lagrangian
 in a 4+1 dimensional space-time
along the fifth dimension (the segment of length $\pi R$)
 gives rise to  a linear moose chiral lagrangian after a suitable
identification of the gauge and link couplings
\cite{Arkani-Hamed:2001nc,Arkani-Hamed:2001ca,Hill:2000mu,Cheng:2001vd}.

For the case of equal couplings $f_i$ we take \be K\to\infty,~~~~
a\to 0,~~~~Ka\to\pi R,\ee where $a$ is the lattice size. We find
\be \epsilon_3\to\frac 1 6\left(\frac g{\tilde g}\right)^2 K.\ee
By introducing the gauge coupling in 5 dimensions by the relation
\be
 {g_5^2}={a}{\tilde g^2},\ee
we get \be \epsilon_3\to \frac 1 6\left(\frac g{g_5}\right)^2 \pi
R,\ee in agreement with Ref. \cite{Foadi:2003xa}.

The discretization of a 5 dimensional gauge theory has been
considered also for the warped metric case
\cite{Randall:2002qr,Falkowski:2002cm}. This  corresponds to a
linear moose with  link couplings given by Eq. (\ref{f-exp}) with
  \be c=\frac{\pi k R}{K}.\ee
This exponential behavior of $f_i$ corresponds to the Randall
Sundrum metric \cite{Randall:1999vf}. Then  $c\to 0$ for
$K\to\infty$ and $a(i-1)\to y$ where $y\in[0,\pi R]$ is the
coordinate along the fifth dimension. That is \be f_i=\bar f
e^{ka(i-1)}\to f(y)=\bar f e^{ky}.\ee Therefore in the warped case
we find, from Eq. (\ref{eq:46}) \be \epsilon_3=\frac {1}{ 4
k}\frac{e^{4k\pi R}-4k\pi R e^{2k\pi R}-1}{(1-e^{2k\pi R})^2}
\left(\frac g{g_5}\right)^2. \ee For large values of $k\pi R$  we
get \be \epsilon_3\to \frac 1 {4k}\left(\frac g{g_5}\right)^2.\ee
Assuming $k\sim M_{Pl}/10$, $R\sim 10^2 M_{Pl}^{-1}$ and
$g_5^2=\pi R g_4^2$, where $g_4$ is the gauge coupling obtained
after dimensional reduction of the fifth dimension, it follows
$\epsilon_3\sim 0.008~ g^2/g_4^2$. In the reflection invariant
case we obtain $\epsilon_3\sim 0.016 ~g^2/ g_4^2$. Therefore, also
in the continuum limit we get a suppression factor although not as
large as in the discrete case.

\section{Further extensions: the planar moose}
\label{section:VI}

 Possible generalizations of the linear moose are obtained   extending the
moose graph in the plane. A realistic model for the electroweak
symmetry breaking must contain  only three independent scalar
fields (the Goldstone bosons necessary to provide the masses to
the electroweak gauge bosons) and will have additional  massive
vector gauge bosons. Then we can immediately show that the only
possible diagrams are the ones with zero loops. In fact a moose
diagram is like a Feynman diagram with lines corresponding to
links and vertices corresponding to gauge groups. Therefore, by
introducing the following notation: \bea E&=& {\rm
number~of~external~
links},\nn\\ I&=& {\rm number~of~internal~ links},\nn\\
V_\ell&=&{\rm number~ of~ gauge~ groups~ with}~ \ell~{\rm
links},\nn\\ L&=&{\rm number ~of ~loops},\nn\\ S&=&{\rm
number~of~remaining~Goldstone~multiplets},\eea we have \be
L=I-(\sum_{\ell}V_\ell -1)\label{loop2},\ee \be
S=I+E-\sum_{\ell}V_\ell\label{loop3}.\ee By using Eqs.
(\ref{loop2}) and (\ref{loop3}) we get \be L=S-(E-1).\ee

In the models considered in this paper we have   associated to the
external links a global symmetry. Therefore we need at least two
external links ($E=2$) in order to get the right weak
phenomenology. This, together with the requirement of one scalar
multiplet ($S=1$), implies that the number of  loops must be equal
to zero.

\begin{figure}[h] \centerline{
\epsfxsize=12cm\epsfbox{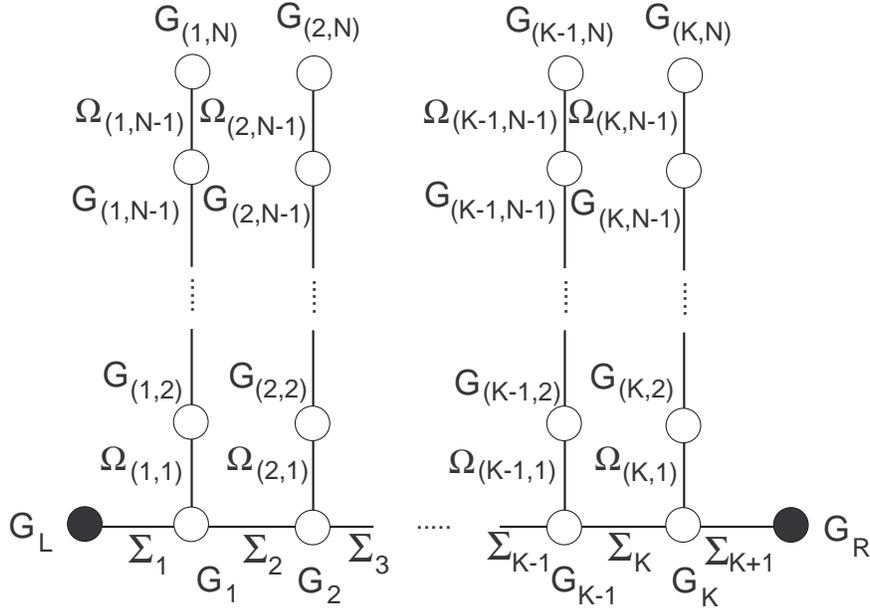} } \caption {{ \it The planar
moose \label{2dim_moose} }}
\end{figure}

 Avoiding loops, the way to generalize the linear moose
to a planar one  is to attach a string to each of the groups $G_i$
as illustrated in Fig. \ref{2dim_moose}.

For simplicity we take all the strings of equal length, with $N-1$
links, but as we shall see, the result can be immediately extended
to strings of different length. As shown in Fig. \ref{2dim_moose}
we introduce gauge groups $G_{(i,j)}$ with $i=1,\cdots,K$ and
$j=1,\cdots, N$, and associated fields $A_{(i,j)}$  with
corresponding gauge couplings $g_{(i,j)}$. However it is
convenient to identify explicitly the gauge fields and couplings
on the original string \be
 A_{(i,1)}=A_i,~~~~g_{(i,1)}=g_i ~~~i=1,\cdots,K.\ee In addition
 to the scalar fields $\Sigma_i$, $i=1,\cdots,K+1$ linking the
 gauge fields $A_i$, we have also  new scalar fields $\Omega_{(i,j)}$,
 $i=1,\cdots,K$, $j=1\cdots, N-1$ linking the gauge fields along
 the vertical direction. We introduce the following notation
 \be
 B_i= g_iA_i -g_{i+1}A_{i+1},~~~~i=0,\cdots,K,\ee with the boundary
 condition \be g_0=g_{K+1}=0.\ee
Notice that the $K+1$ fields   $B_i$ are not independent, since
\be \sum_{i=0}^K B_i= 0.\label{independence}\ee We define also
 \be
 V_{(i,j)}=g_{(i,j)}A_{(i,j)}-g_{(i,j+1)}A_{(i,j+1)},~~~~i=1,\cdots,
 K,~~j=1,\cdots,N-1.\ee
Then the vector boson mass term will be given by \be{\cal
L}_{mass}=\frac 1 2\sum_{i=1}^{K+1} f_i^2 B_{i-1}^2+\frac 1
2\sum_{i=1}^K\sum_{j=1}^{N-1}h_{(i,j)}^2 V_{(i,j)}^2,
\label{mass_planar}\ee where $h_{(i,j)}$ are new $K\times (N-1)$
link couplings. As shown in Appendix A the result for $\epsilon_3$
is \be \epsilon_3=g^2\sum_{i=1}^K\frac{y_i(1-y_i)}{\tilde
g_i^2},\ee where \be \frac 1{\tilde g_i^2}=\sum_{j=1}^N \frac 1
{g_{(i,j)}^2}.\label{new_gau}\ee We see that the only effect of
attaching a string at any of the initial groups $G_i$ is simply to
define a new gauge coupling according to Eq. (\ref{new_gau}).

As in the linear moose we can consider the continuum limit. Let us
introduce a lattice size $b$ along the vertical direction and take
the limit \be
 b\to 0,~~~~N\to\infty,~~~~bN\to \pi R^\prime.\ee
By defining a
 six-dimensional gauge coupling  as
 \be
 \frac {ab}{g_6^2}=\frac 1{\tilde g^2},\ee
  we get
 \be
 \epsilon_3=\frac 1 6 \left(\frac g{g_6}\right)^2 \pi^2  R R^\prime,\ee
 for the constant case $f_i=f$, while
for exponential $f_i$ (see Eq. (\ref{f-exp})) \be \epsilon_3=\frac
1 4\left(\frac g{g_6}\right)^2\frac{\pi R^\prime}{k}. \ee
 Notice that we do not get a suppression
 factor from the vertical links, since the result for $\epsilon_3$
 does not depend on these variables.

\section{Conclusions}
\label{conclusions}

Models with replicas of gauge groups have been recently considered
because they  appear in the deconstruction of five dimensional
gauge models which have been used  to describe the electroweak
breaking without the Higgs
\cite{Csaki:2003dt,Csaki:2003zu,Nomura:2003du,Barbieri:2003pr,Foadi:2003xa}.
The four dimensional description is based on the linear moose
lagrangians that were already proposed in technicolor and
composite Higgs models \cite{Georgi:1986hf}. In general these
models satisfy the constraints arising
 from the parameters $T$ and $U$ (or $\epsilon_1$ and $\epsilon_2$) due to
the presence of a custodial $SU(2)$ symmetry. However such models
generally  give a correction of order ${\cal O}(1)$ (${\cal
O}(10^{-2}$)) to the parameter $S$ ($\epsilon_3$). In this paper
we have considered a linear moose based on replicas of $SU(2)$
gauge groups and with the electroweak gauge groups $SU(2)_L$ and
$U(1)_Y$ at the two ends of the moose string. After having
obtained
 a general expression for the parameter $\epsilon_3$,
we have shown that a unique solution exists which guarantees
$\epsilon_3=0$. The corresponding model has  an additional
custodial symmetry which protects this result. It is obtained from
the  linear moose when a  link is cut and  a non local field
connecting the two ends of the moose is included. This solution is
a generalization of a simplest case, corresponding to the
degenerate BESS model. It contains additional vector resonances,
however their contribution to the $S$ parameter is zero at the
leading order in the electroweak couplings. At the same order the
new resonances do not couple to the longitudinal $W$ and $Z$. As a
consequence the breaking of partial wave unitarity is expected to
happen at the same scale as in the Higgless SM and it is not
postponed to higher scales.

We have also shown that it is possible to control the size of
$\epsilon_3$ by taking one of the link couplings much smaller than
the other ones.

A generalization to the planar case has been also investigated: we
have shown that no loops are allowed in the moose graph and that
with a convenient redefinition of the gauge couplings the result
for $S$ is the same as for the linear moose case.

\acknowledgements

We would like to thank  Massimiliano Grazzini for his help in a
first phase of this paper and for a critical reading of the
manuscript. D.D. would like to thank Riccardo Rattazzi for
stimulating discussions and CERN Theory Division for hospitality.

\appendix

\section{Evaluation of $\epsilon_3$}
\label{appendixA} In this Appendix we will obtain explicit
expressions for  $\epsilon_3$ both for the linear moose and for
its planar generalization considered in Section \ref{section:VI}.
We start  evaluating the inverse of the vector boson mass matrix,
$M_2$. Actually, for $\epsilon_3$ we need only the elements
$(M_2^{-1})_{1i}$ and $(M_2^{-1})_{Ki}$ where the index $i$ runs
over all the gauge fields. A technique to determine $M_2^{-1}$ is
to  add to the mass term a source, \be{\cal L}_{mass+source}=\frac
1 2 A^TM_2A-JA.\ee Evaluating from this lagrangian the equations
of motion and solving for the $A$'s in terms of the sources $J$'s
we find \be A=M_2^{-1}J,\label{eqs_motion}\ee from which we can
read the relevant matrix elements.

\subsection{The linear moose}

The mass term in Eq. (\ref{lmass}) can be written as \be {\cal
L}_{mass}=\frac 1 2\sum_{i=1}^{K+1}f_i^2 B_{i-1}^2.\ee The
variables $B_i$ are the analogue of canonical momenta in the
discrete case and are given by \be
 B_i= g_iA_i -g_{i+1}A_{i+1},~~~~i=0,\cdots,K,\ee with the boundary
 condition $g_0=g_{K+1}=0$.
Notice that the $K+1$ fields   $B_i$ are not independent, since
\be \sum_{i=0}^K B_i= 0.\label{indep}\ee

 Therefore
we solve the equations of motion (\ref{eqs_motion}), which involve
three nearest neighborhoods by solving first in the $B_i$'s and
then inverting the relation between the $B_i$'s and the fields
$A_i$. These equations involve only first neighborhoods. This is
the analogue of converting a second order differential equation in
a pair of first order equations.

The equations of motion can be written in the following form
\be-f_i^2B_{i-1}+f_{i+1}^2B_i=L_i,~~~~i=1,\cdots,K,\ee where we
have redefined the sources as \be L_i=\frac{J_i}{g_i}.\ee We can
solve for all the $B_i$, $i=1,\cdots,K$ in terms of $B_0$ finding
\be B_i=\frac 1{f_{i+1}^2}\left(\sum_{j=1}^i L_j+f_1^2
B_0\right),~~~~~i=1,\cdots,K.\label{eq:B7}\ee  It is convenient to
introduce the following variables \be \frac
1{f^2}=\sum_{i=1}^{K+1}\frac 1
{f_i^2},~~~~x_i=\frac{f^2}{f_i^2},~~~~i=1,\cdots,K+1,\ee \be
y_i=\sum_{j=1}^i x_j,~~~z_i=\sum_{j=i+1}^{K+1}x_j,\ee with  the
properties \be y_i+z_i=1,~~~~y_1=x_1,~~~~z_K=x_{K+1}.\ee

By summing the Eqs. (\ref{eq:B7})
 over $i$ from 1 to $K$ and using Eq. (\ref{indep}) we get a relation for
$B_0$ which can be easily solved obtaining \be
B_0=-\frac{x_1}{f^2}\sum_{i=1}^K z_i\,L_i,\ee and \be
B_i=\frac{x_{i+1}}{f^2}\left(\sum_{j=1}^i\,y_j\,L_j-\sum_{j=i+1}^K\,z_j\,L_j\right).\ee
By using  the discrete step function given in Eq. (\ref{eq:25}) we
can write \be
B_i=\frac{x_{i+1}}{f^2}\sum_{j=1}^K\left(\theta_{i,j}y_j-\theta_{j,i+1}z_j
\right)L_j,~~~i=0,\cdots,K\label{momenta}.\ee Notice that this
equation holds also for $i=0$, due to the properties of the
discrete $\theta$-function. Further we need to reexpress the
fields $A_i$ in terms of the $B_i$'s. We find \be A_i=\frac 1
{g_i}\sum_{j=1}^K\theta_{j,i}B_j.\ee Using Eq. (\ref{momenta}) we
obtain
\be(M_2^{-1})_{1i}=\frac{x_1z_i}{g_1g_if^2},~~~~(M_2^{-1})_{iK}=
\frac{x_{K+1}y_i}{g_{K}g_if^2}.\ee Therefore, from the expression
(\ref{eps3}) we get \be
\epsilon_3=g^2\sum_{i=1}^K\frac{z_iy_i}{g_i^2}=g^2\sum_{i=1}^K\frac{(1-y_i)y_i}{g_i^2}.\ee

\subsection{The planar moose}

Starting from the mass term of the planar case, Eq.
(\ref{mass_planar}), we get the following set of equations of
motion by differentiating with respect to $A_{(i,j)}$ \be -f_i^2
B_{i-1}+f_{i+1}^2  B_i+ h_{(i,1)}^2
V_{(i,1)}=\frac{J_i}{g_i}\equiv
L_i,~~~i=1,\cdots,K,\label{horizontal} \ee \be
-h^2_{(i,j-1)}V_{(i,j-1)}+h^2_{(i,j)}V_{(i,j)}=\frac{J_{(i,j)}}{g_{(i,j)}}\equiv
L_{(i,j)},~~~i=1,\cdots,K,~~~j=2,\cdots,N.\label{vertical}\ee It
is also convenient to introduce \be L_{(i,1)}=L_i.\ee The solution
of Eq. (\ref{vertical}) is \be V_{(i,j)}=-\frac
1{h_{(i,j)}^2}\sum_{m=j+1}^NL_{(i,m)}.\ee Inserting this result
inside Eq. (\ref{horizontal}) we obtain \be -f_i^2
B_{i-1}+f_{i+1}^2  B_i=\tilde L_i,\ee with \be\tilde
L_i=\sum_{j=1}^NL_{(i,j)}=\sum_{j=1}^N\frac{J_{(i,j)}}{g_{(i,j)}}.\ee
These equations are the same as for the linear moose with the
substitution $L_i\to\tilde L_i$. Therefore we get immediately \be
(M_2^{-1})_{1,(i,j)}=\frac{x_1 z_i}{f^2 g_1g_{(i,j)}},~~~~
(M_2^{-1})_{K,(i,j)}=\frac{x_{K+1} y_i}{f^2 g_Kg_{(i,j)}},\ee
where the variables $x_i$, $y_i$, $z_i$ and $f^2$ are the same as
for the linear moose. Therefore the result is \be
\epsilon_3=g^2\sum_{i=1}^K\frac{y_i(1-y_i)}{\tilde g_i^2},\ee
where \be \frac 1{\tilde g_i^2}=\sum_{j=1}^N \frac 1
{g_{(i,j)}^2}.\label{new_gauge}\ee

\section{Solutions to $\epsilon_3=0$}
\label{appendixB}

We want to prove the following statement: {\it $\epsilon_3=0$ if
and only if one or more $f_i$'s are sent to zero  in an
independent way}.

We start noticing that due to the condition $\sum_{i=1}^{K+1}
x_i=1$,  all the $y_i$'s, defined in Eq. (\ref{xi}), are such that
$0\le y_i\le 1$. Since $\epsilon_3$ is made off of positive terms,
each of them proportional to $0\le y_i(1-y_i)\le 1/4$, in order to
get a vanishing $\epsilon_3$ we need $y_i(1-y_i)=0$ for all values
of $i$. This implies $y_i=0$ or $y_i=1$ for all $i$'s. However,
since $x_i=y_{i}-y_{i-1}$, the same properties must hold true for
the quantities $x_i$, that is $x_i=0$ and $x_i=1$ for all values
of $i$. Also we have already shown that $\epsilon_3=0$ if one of
the $f_i$'s is chosen to vanish, see Eq. (\ref{step}). Therefore
if we send to zero several link couplings $f_i$, we get
$\epsilon_3=0$ {\it a fortiori}.

Let us now show that if we send more than one  $f_i$ to zero in a
correlated way, then $\epsilon_3\not=0$. Assume that $p$ of the
constants $f_i$ go to zero with the variable $\eta$ in a
simultaneous way \be f_i^2=c_i\eta,~~~c_i>0,~~~~i\in {\cal P}.\ee
Here $i$ takes $p$ values in the subset ${\cal P}$ of the set
$(1,\cdots,K+1)$. Notice that the assumption of correlation
 implies that the coefficients $c_i$ are strictly
positive. In the limit $\eta\to 0$ we get immediately \be\frac
1{f^2}\approx\frac 1\eta\sum_{i\in{\cal P}}\frac 1{c_i}.\ee It
follows \be x_i=\left\{ \begin{array}{l}{\dd{\frac 1{c_i}\frac
1{\sum_{j\in{\cal P}}\frac 1{c_j}}~~~i\in{\cal
P},}}\\{{\dd{0~~~~~~~~~~~~~~~i\not\in{\cal P}.}}}\\
\end{array}\right. \ee
Unless the set ${\cal P}$ contains a single element we have \be
x_i<1,~~~i\in{\cal P}.\ee As a consequence some of the $y_i$'s is
neither zero nor one and $\epsilon_3$ is not vanishing.

%\bibliographystyle{apsrev}
%\bibliography{moose}

\end{document}